# THE GENETIC CODE INVARIANCE: WHEN EULER AND FIBONACCI MEET

## Tidjani Négadi

*Address:* Department of Physics, Faculty of Science, University of Oran, 31100, Es-Sénia, Oran, Algeria, E-mail: tnegadi@gmail.com.

**Abstract**: *The number of atoms in the four ribonucleotides uridine monophosphate (UMP), cytidine monophosphate (CMP), adenine monophosphate(AMP) and guanine monophosphate (GMP) is taken as a key parameter. A mathematical relation describing the condensation of the three basic sub-units a nucleobase, a ribose and a phosphate group, to form a ribonucleotide, is first obtained from this parameter. Next, the use of the latter and Euler's totient function is shown to lead to the atom number content of the 64 codons and also to Rakočević's pattern "2×3456". Finally, selected finite sums of Fibonacci numbers are shown to lead to the nucleon number content of the amino acids in various degeneracy patterns, and also to the multiplet structure of the 20 amino acids as well as to the degeneracy.*

## 1. INTRODUCTION

This contribution to Symmetry Festival 2013 is devoted to new and original applications of some well-known arithmetic functions as famous Euler's function $\varphi(n)$, and also, no less famous Fibonacci's series, to the study of the genetic code structure. Since its deciphering in the sixties of the last century (Nirenberg et al., 1965), the genetic code, *i.e.*, the mapping between 64 codons and 20 amino acids (see Table 1), has been shown to exhibit many remarkable arithmetical regularities with respect to its chemical content: atom numbers (Rakočević, 2006, 2009, 2011, 2013, Négadi, 2012), nucleon numbers (shCherbak, 1994, 2003; Downes and Richardson, 2002) and also carbon atom numbers (Yang, 2004, Négadi, 2005). The present work is a continuation of these investigations but, this time, we shall start not from the genetic code itself, as it is, but from the basic molecule of life RNA, more exactly from its four units the ribonucleotides *uridine monophosphate* (UMP), *cytidine monophosphate* (CMP), *adenosine monophosphate* (AMP) and *guanosine monophosphate* (GMP). It is thought that these four molecules may have played a key role in the early evolution of life (Gilbert's RNA-World hypothesis, Gilbert, 1986). These four basic molecules have the following molecular brut formulae and, in parenthesis, those for the nucleobases (http://www.ncbi.nlm.nih.gov):

UMP: $C_9H_{13}N_2O_9P$  (U: $C_4H_4N_2O_2$)

CMP: $C_9H_{14}N_3O_8P$  (C: $C_4H_5N_3O_1$)

AMP: $C_{10}H_{14}N_5O_7P$  (A: $C_5H_5N_5$)

GMP: $C_{10}H_{14}N_5O_8P$  (G: $C_5H_5N_5O_1$)

| UUU F | UUC F | UCU S | UCC S | CUU L | CUC L | CCU P | CCC P |
|---|---|---|---|---|---|---|---|
| UUA L | UUG L | UCA S | UCG S | CUA L | CUG L | CCA P | CCG P |
| UAU Y | UAC Y | UGU C | UGC C | CAU H | CAC H | CGU R | CGC R |
| UAA stop | UAG stop | UGA stop | UGG W | CAA Q | CAG Q | CGA R | CGG R |
| AUU I | AUC I | ACU T | ACC T | GUU V | GUC V | GCU A | GCC A |
| AUA I | AUG M | ACA T | ACG T | GUA V | GUG V | GCA A | GCG A |
| AAU N | AAC N | AGU S | AGC S | GAU D | GAC D | GGU G | GGC G |
| AAA K | AAG K | AGA R | AGG R | GAA E | GAG E | GGA G | GGG G |

**Table 1**: The standard genetic code table

Importantly, throughout this work, we shall consider, the total *atom-number* in these four molecules as a *key* parameter derived from the above formulae and to be equal to 144. As a link to the second section, it is interesting to note that the number of hydrogen atoms is equal to 55 (the 10$^{th}$ Fibonacci number) and the total number of the remaining atoms (carbon, nitrogen, oxygen and phosphorus) is equal to 89 (the 11$^{th}$ Fibonacci number), therefore leading to the 12$^{th}$ Fibonacci number: 144=55+89. In the first section, using Euler's $\varphi$-function and some other elementary arithmetic functions, we derive, from the total number of atoms in the four ribonucleotides (144), several interesting outcomes concerning the genetic code mathematical structure and its relation to the 20 aminoacyl-tRNA synthetases that are known to form the genetic code. In addition, we derived an *invariance equation* for the total number of atoms in the 64 codons. In the second section, as a starting point, some few interesting finite sums of Fibonacci numbers are considered, from which several other meaningful and complementary results are obtained. For example, several degeneracy-patterns for the total *nucleon number* in the 61 coded amino acids and the multiplet pattern of the 20 amino acids, as well as their codon degeneracy, are obtained. Besides Euler's $\varphi$ function and the Fibonacci series, the mathematical tools used in this paper, we also utilized the arithmetic function $B_0(n)$, introduced by Négadi in his earlier work (Négadi, 2007, 2008, 2009, 2011). It is defined as the sum of (i) the arithmetic function $a_0(n)$ giving the sum of the prime factors of an integer n (from the Fundamental Theorem of Arithmetic) including the multiplicity, (ii) the sum of the prime indices of the prime factors, noted here SPI($n$), and, finally the number of the prime factors, the so called Big Omega function, $\Omega(n)$. For example for $12=2^2\times 3$, we have $B_0(12)=a_0(12)+\text{SPI}(12)+\Omega(12)=$ $(2+2+3)+(1\times 1+2)+(2+1)=7+4+3=14$.

## 2. EULER'S TOTIENT FUNCTION AND ATOM NUMBERS

Euler's totient function, or $\varphi$-function, is a very important function in mathematics and has some applications as, for example, in public key cryptography or in graph theory. Here, we are demonstrating that it has also interesting applications in mathematical biology. The $\varphi$-function of a positive integer n greater than 1 is defined to be number of positive integers less than n that are coprime to $n$ (with $\varphi(1)=1$). This function is easily computable, using Maple for example. First, we begin by applying this function to our key number 144, the total number of atoms in the four ribonucleotides. We get

$$\varphi(144)=48 \tag{1}$$

The number 144 has therefore 48 coprimes. Examining the genetic code table (Table 1), we have concluded that each of the four nucleobases U, C, A and G (or ribonucleotides UMP, CMP, AMP and GMP) appears exactly 48 times in the table. Next, we form the following sum (using Eq.(1))

$$144+\varphi(144)=192 \tag{2}$$

In addition, we obtained an interesting result in terms of the total number of nucleobases in the table (48×4=192). It is also possible to compute the *sum* of all the coprimes of a number *n* using the following formula (from a well known theorem)

$$\frac{1}{2}[n\times\varphi(n)] \tag{3}$$

Applying this formula to our case gives

$$\frac{1}{2}[144\times\varphi(144)] = 3456 \tag{4}$$

The number 3456, is also very interesting number, as we will demonstrate later but, for now on, we rearrange the above relation to obtain, from Eqs,(1) and (4)

$$\varphi(144)\times144=48\times144=2\times3456=6912 \tag{5}$$

The number in Eq.(5), rewritten as 48×(34+35+37+38)=48×144=6912, gives us exactly the *total number of atoms in the 64 RNA-codons* (using the ribonucleotides UMP, CMP, AMP and GMP, cf., the chemical formulae in the introduction) in a form that interestingly fits, in particular, to Rakocevic's "2×3456" pattern, Petoukhov's even codons/odd codons that follow a "2×3456" pattern too, and, with some further computation, the atom number pattern of the codons for the 20 amino acids classified according to their Class-I/Class-II aminoacyl-tRNA synthetases (see below). Before examining these matters, let us consider again the number of atoms in the four ribonucleotides and compute its $B_0$-function (cf., the introduction). We get ($144=2^4\times3^2$)

$$B_0(144)=a_0(144)+SPI(144)+\Omega(144)=14+8+6=28 \tag{6}$$

Grouping the numbers in the right of Eqs,(6) in *two* ways, 22+6=20+8=28, we obtain

$$22=20+8-6 \tag{7}$$

This is part of the "condensation" equation for the formation of a ribonucleotide (see Rakočević, 1997, for a calculation). As a matter of fact, a ribonucleotide is made of three sub-units: a phosphate group ($H_3PO_4$, 8 atoms), a ribose ($C_5H_{10}O_5$, 20 atoms) and a nucleobase N. Note, that the phosphate group and the ribose are *identical* in any given set of nucleobases (so representing a kind of regularity). Also, in the condensation process, *two water molecules are released*. Adding a nucleobase N, more exactly its number of atoms which we represent by the same letter, to both sides of Eq.(7), we have

$$N+\text{ribose}+\text{phosphate} - (2\ H_2O) = N+20+8-6 = N+22 \tag{8}$$

Here, figure 6 represents the number of atoms in two water molecules ($2H_2O \to 2\times 3$). In this way, we could obtain the number of atoms in a given ribonucleotide by simply adding 22 atoms to those of the corresponding nucleobase (for example UMP: 12+22=34). Therefore, *the ribonucleotides appear to carry with them, in their atom number content, the imprint of their construction from their sub-units, the ribose, the phosphate group and the nucleobase*. Now, we introduce the following new function $\Phi(n) \equiv \varphi(\sigma(n))$ of an integer $n$, where $\sigma$ is the sum of the divisor's function, and apply it to the total number of atoms in 64 codons, given in Eq.(5). We get

$$\Phi(6912) = 6912 \tag{9}$$

One can see that this latter number remains *unchanged*. In fact, this number is the 8$^{\text{th}}$ member[1] of a mathematical sequence of numbers with the property $\varphi(\sigma(n))=n$. It appears therefore that Eq.(9) could constitute an "*invariance equation*" for the 64-codons of the genetic code table, for their total number of atoms is conserved. We could also infer from the number 144 and its number of coprimes 48, from Eq.(1), other important quantities connected with the amino acids and the structure of the genetic code. Below, we give somea few of them. First, we have

$$\varphi(144)+\varphi(48) = 61+3 = 64, \tag{10}$$

where we have used the fact that among the 16 coprimes of the number 48 (**1**, 5, 7, 11, 13, 17, 19, 23, **25**, 29, 31, **35**, 37, 41, 47) 13 are prime and 3 non-prime (1, 25 and 35), so that by introducing the partition 16=13+3 in Eq.(10) we get: (48+13)+3=61+3=64. This relation gives us the total number of codons in the genetic code with 61 amino acids coding codons and 3 stop codons. At second, applying Eq.(3), giving the sum of the coprimes, to the number 48 itself gives

$$\frac{1}{2}[48\times\varphi(48)] = 384 \tag{11}$$

---

[1] 1, 2, 8, 12, 128, 240, 720, **6912**, 32768, 142560, 712800, 1140480, 1190400, 3345408, 3571200, 5702400, … (Some 365 terms of this sequence are known but it is an open problem whether it is infinite or not, Ref. Guy R. K. (2004) *Unsolved Problems in Number Theory*, New York: Springer-Verlag: B42; see also http://oeis.org/A001229).

The number in Eq.(11) is the same as the total number of atoms in the 20 amino acids, *side-chains and blocks* (see for example Négadi, 2009, and Table 3). Third, we compute the sum of all the divisors of 144 and obtain

$$\sigma(144) = 403 \qquad (12)$$

This last number fits the total number of atoms in the *biosynthetic precursors* for 23 amino acids[2] comprising 8 four-codons and 15 non-four codons (Rakočević, 1998; see also Négadi, 2011). We shall return to this last result, in detail, below. Now, we return to Eq.(5), in connection with the "2×3456" pattern. This latter seems present in several works on the genetic code. It has been first mentioned by Rakočević (2011). He showed that there are 3456 atoms within the codons in two inner columns and 3456 atoms within the codons in two outer columns in his GCT (Genetic Code Table). In the above mentioned work, entitled *Genetic Code: four diversity types of protein amino acids*, he established, in a special section entitled "The uniqueness of number 26 through unique pair 25-36" the following interesting algebraic system of equations with the solutions in parenthesis:

$$x_1+y_1=36=\mathbf{6}^2 \quad (x_1=26; y_1=10)$$
$$x_2+y_2=25=\mathbf{5}^2 \quad (x_2=17; y_2=8) \qquad (R)$$
$$x_1-y_1=16=\mathbf{4}^2$$
$$x_2-y_2=09=\mathbf{3}^2$$

what he linked to arithmetical regularities in specific arrangement of natural numbers in the decimal numbering system with Shcherbak's "simulation" analogs. When dividing the set of the 20 amino acids into four "diversity types" (here $x_i$ and $y_i$, i=1,2, are natural numbers and will be connected to codon numbers, see below). These four diversity types are the following (i) two amino acids (G, P) both without standard hydrocarbon side chains, (ii) four amino acids [(A, L), (V, I)], all with standard hydrocarbon side chains, (iii) six amino acids, as three pairs [(F, Y), (H, W), (C, M)] two aromatic, two hetero aromatic and two "hetero" non-aromatic and finally (iv) eight amino acids as four pairs [(S, T), (D, E), (N, Q), (K, R)] all with a functional group (see the details in Rakočević, 2011). He also noted a striking "correspondence" between the number 3456 and the exponents (3, 4, 5, 6), shown in the Eq.(R) above, in bold. It is interesting that we could derive, here, the pattern "2×3456" and also the invariance property found in Eq.(9), using matrices from our matrix method introduced a few years ago (see Négadi, 2001). We could compute the following 8×8 codon atom number matrix (using Kronecker products, see Négadi, 2001):

$$C = \begin{pmatrix} 102 & 103 & \mathbf{103} & \mathbf{104} & \mathbf{103} & \mathbf{104} & 104 & 105 \\ 105 & 106 & \mathbf{106} & \mathbf{107} & \mathbf{106} & \mathbf{107} & 107 & 108 \\ \mathbf{105} & \mathbf{106} & \mathbf{106} & \mathbf{107} & \mathbf{106} & \mathbf{107} & \mathbf{107} & \mathbf{108} \\ \mathbf{108} & \mathbf{109} & \mathbf{109} & \mathbf{110} & \mathbf{109} & \mathbf{110} & \mathbf{110} & \mathbf{111} \\ \mathbf{105} & \mathbf{106} & \mathbf{106} & \mathbf{107} & \mathbf{106} & \mathbf{107} & \mathbf{107} & \mathbf{108} \\ \mathbf{108} & \mathbf{109} & \mathbf{109} & \mathbf{110} & \mathbf{109} & \mathbf{110} & \mathbf{110} & \mathbf{111} \\ 108 & 109 & \mathbf{109} & \mathbf{110} & \mathbf{109} & \mathbf{110} & 110 & 111 \\ 111 & 112 & \mathbf{112} & \mathbf{113} & \mathbf{112} & \mathbf{113} & 113 & 114 \end{pmatrix}$$

---

[2] The biosynthetic precursors of the amino acids are: 1. *Ribose-5-phosphate* (25; H); 2. *3-Phosphoglycerate* (18; G, **S**, C); 3. *Pyruvate* (10; A, **L**, V); 4. *2-Oxoglutarate* (16; P, E, Q, **R**); 5. *Phosphoenolpyruvate+Eritrose-4-phosphate* (36; W, F, Y); 6. *Oxaloacetate* (13; T, M, I, D, N, K). In the parenthesis, the number of atoms in the precursor is followed by the corresponding amino acids (the sextets are indicated in bold character).

It is easy to recognise in the above atom number matrix (where each matrix element corresponds to a codon in Table 1 at the same location) that the total number of atoms in the *colums* **3**, **4**, **5** and **6** is 3456 and the total number of atoms in the *rows* **3**, **4**, **5** and **6** is also 3456. This is the same result as Rakočević's one, *i.e.*, the pattern "2×3456". This pattern could also be derived from the above matrix by taking the product of its dimension and its trace:

$$\text{Dim}(C) \times \text{Tr}(C) = 8 \times 864 = 6912 \qquad (13)$$

Some numbers in the diagonal show multiplicity. By grouping them in the trace – the non repeating numbers, on the one hand, and the repeating ones, on the other – we get finally

$$[102+106+110+114] \times \text{Dim}(C) + (2 \times 106 + 2 \times 110) \times \text{Dim}(C) = 3456 + 3456 \qquad (14)$$

which is again the pattern "2×3456". By noting that the dimension and the trace are both *invariant* objects and so does their product, we find the invariance property found in Eq.(9) in another way. Concerning the numbers in Rakočević's system in Eq.(R), we could recover them as follows. Let us take the prime factor decomposition of the sum of divisors function of the number 3456: $\sigma(3456)=10200=2^3 \times 3 \times 5^2 \times 17$ and compute its $B_0$-function (see the introduction). Using the additivity property of the $B_0$-function (also called integer logarithm), we have $B_0(2^3 \times 3 \times 5^2 \times 17) = B_0(2^3 \times 3 \times 5^2) + B_0(17) = 36+25 = 61$, which is the result shown in Eq.(R) above with $B_0(2^3 \times 3 \times 5^2) = x_1+y_1$ and $B_0(17) = x_2+y_2$. We could also write, by regrouping in a particular way the different components of the $B_0$ functions,

$$B_0(2^3\times3\times5^2)=a_0(5^2)+[B_0(2^3\times3)+SPI(5^2)+\Omega(5^2)]=10+[18+6+2]=10+26=36 \quad (15)$$

$$B_0(17)=a_0(17)+[SPI(17)+\Omega(17)]=17+[7+1]=17+8=25 \quad (15)'$$

which seemingly fit to Rakočević's solutions of the above system of equations and are shown in the parenthesis in Eq.(R). We have thus, fully, recovered the solutions of Rakočević's system describing the four diversity types: (i) (G, P) 8 codons and (ii) [(A, L), (V, I)] 17 codons, on the one hand, and (iii) [(F, Y), (H, W), (C, M)] 10 codons and [(S, T), (D, E), (N, Q), (K, R)] 26 codons, on the other, as clustered by Rakočević. Let us say also some words concerning the pattern "2×3456" with respect to the work by Petoukhov on the genetic code and relying on data by Perez (2010) on the human genome. Petoukhov (2012) showed that "the structure of the whole human genome is connected with the equal division of the whole set of 64 triplets into the even-set of 32 triplets and the odd-set of 32 triplets" and that the frequencies of the codons in the two sets are almost nearly equal (~0.12%), that is, a very near balance. By turning, here, to the RNA-version of his list of codons (with the usual transcription DNA→mRNA rules T→A, C→G, A→U, G→C) we have for the even set: {GGG, GGU, GUG, GUU, UGG, UGU, UUG, UUU, GAG, GAU, GCG, GCU, UAG, UAU, UCG, UCU, AGG, AGU, AUG, AUU, CGG, CGU, CUG, CUU, AAG, AAU, ACG, ACU, CAG, CAU, CCG, CCU} and for the odd set: {GGA, GGC, GUA, GUC, UGA, UGC, UUA, UUC, GAA, GAC, GCA, GCC, UAA, UAC, UCA, UCC, AGA, AGC, AUA, AUC, CGA, CGC, CUA, CUC, AAA, AAC, ACA, ACC, CAA, CAC, CCA, CCC}. In the even set, one counts 32 U, 16 C, 16 A and 32 G. In the odd set, we have 16 U, 32 C, 32 A and 16 G. Using the number of atoms in the ribonucleotides, mentioned above, we get 34×32+35×16+37×16+38×32=3456 atoms for the even set and 34×16+35×32+37×32+38×16=3456 atoms for the odd set. Therefore, we found therefore here also a "2×3456" pattern, i.e., an exact balance.

To close this section, we consider the 20 Aminoacyl t-RNA synthetases (AARSs) which are known to establish the genetic code biochemically. In this case, also, we are going to show the presence of the "2×3456" pattern. The 20 AARSs are divided in two classes of 10 members in each, see Table 2 below.

(In Table 2,

- columns II give the distribution of the (ribo) nucleotides in the codons of a given amino acid, including the degeneracy;
- columns III give the number of nucleons in a given amino acid, including the degeneracy;

- columns IV give the number of atoms in the codons of a given amino acid, using the nucleobases and including the degeneracy;
- columns V give the number of atoms in the codons of a given amino acid, using the ribonucleotides UMP, CMP, AMP and GMP and including the degeneracy.
  The last row gives the sums.)

Rakočević found atom number balances within the 20 amino acids directed by these AARSs (see for example Rakočević, 2011) and, it is interesting for us here that he showed: there are exactly 3456 atoms in the 32 codons for the 10 amino acids in the AARS class-II, using UMP, CMP, AMP and GMP (Rakočević, 2006), (see also Table 2). Below, we shall find how the other "3456" part is structured.

| | Class-I AARSs | | | | | Class-II AARSs | | | |
|---|---|---|---|---|---|---|---|---|---|
| aa | II | III | IV | V | aa | II | III | IV | V |
| Y | 3,1,2,0 | 214 | 79 | 211 | F | 5,1,0,0 | 182 | 73 | 205 |
| W | 1,0,0,2 | 130 | 44 | 110 | N | 1,1,4,0 | 116 | 85 | 217 |
| C | 3,1,0,2 | 94 | 81 | 213 | H | 1,3,2,0 | 162 | 81 | 213 |
| Q | 0,2,3,1 | 144 | 87 | 219 | K | 0,0,5,1 | 144 | 91 | 223 |
| L | 9,5,2,2 | 342 | 235 | 631 | D | 1,1,2,2 | 118 | 87 | 219 |
| R | 1,5,4,8 | 600 | 265 | 661 | S | 6,6,3,3 | 186 | 243 | 639 |
| M | 1,0,1,1 | 75 | 43 | 109 | P | 1,9,1,1 | 164 | 160 | 424 |
| I | 4,1,4,0 | 171 | 121 | 319 | T | 1,5,5,1 | 180 | 168 | 432 |
| E | 0,0,3,3 | 146 | 93 | 225 | G | 1,1,1,9 | 4 | 184 | 448 |
| V | 5,1,1,5 | 172 | 168 | 432 | A | 1,5,1,5 | 60 | 172 | 436 |
| | | 2088 | 1216 | 3130 | | | 1316 | 1344 | 3456 |

**Table 2**: The two classes of Aminoacyl-tRNA synthetases (see the text)

We already know that the number 3456 is obtained as the sum of the coprimes of the total number of atoms in the four ribonucleotides, 144, see Eq.(4), and we have therefore, given, the total number of atoms in the 32 codons for the class-II AARS. The missing total number of atoms in the remaining 29 codons for class-I AARS will be found as follows. Let us consider again the function $\Phi(n) \equiv \varphi(\sigma(n))$ introduced earlier in Eq.(9) and apply it, *iteratively*, starting from the number 3456. We obtain:

1  $N_1 = 3456$
2  $N_2 = \Phi(3456) = 2560$
3  $N_3 = \Phi(2560) = 1800$
4  $N_4 = \Phi(1800) = 2880$
5  $N_5 = \Phi(2880) = 3024$
6  $N_6 = \Phi(3024) = 3840$
7  $N_7 = \Phi(3840) = 3456$

First, we see that there is a cyclicity of order six: after 6 iterations, we recover the number 3456 ($N_7 = N_1$). Second, at the first iteration, we have the number 2560. It appears that there are 2560 atoms in the 61 codons not using the ribonucleotides UMP, CMP, AMP and GMP, but the nucleobases U, C, A and G (see Table 2, column IV: 1216+1344=2560). Third, the four remaining numbers, ranked 3, 4, 5 and 6: 1800, 2880, 3024, 3840, which *a-priori* have no special meaning have a *mean* equal to 2886. This last number is equal to the sum of the number of atoms in the 61 (coding) codons, computed using the nucleobases, namely 2560 (see above), and 326 – the number of atoms in the three stop codons (UAA, UAG, UGA) computed using the ribonucleotides UMP, CMP, AMP and GMP: $3 \times 34 + 4 \times 37 + 2 \times 38 = 326$, as computed by Rakočević (Rakočević, 2009). Finally, for the difference 2886-2560=326 is as legitimate as its inverse 2560-2886= -326, we introduce both of them in the sum $N_7 + N_1 = 6912$ to obtain

$$3456 - 326 + 326 + 3456 = (3130 + 326) + 3456 = 3456 + 3456 = 6912 \qquad (16)$$

In the relation found above, 3130 is the number of atoms in the 29 codons for the amino acids in the class-I AARS (see Table 2). Also, the three stop codons UAA, UAG, UGA share their two first bases with amino acids in the class-I AARSs (Y and W) so it is natural to "count" them in this latter AARS class. In this way, we get again the "2×3456" pattern (see Eq.(16). Here, we used UMP, CMP, AMP and GMP in all the calculations. We have thus obtained the correct number of atoms in the codons of class-I AARSs (3130), the number of atoms in the codons of class-II AARss (3456) as well as the number of atoms in the three stop codons (326).

## 3. FIBONACCI NUMBERS AND NUCLEON NUMBERS

In this section, we consider the Fibonacci sequence and show that it could be fruitfully used to compute (a) the number of *nucleons* in the amino acids, in various characteristic patterns, (b) the multiplet structure of the 20 amino acids, as well as (c) the degeneracy. This sequence has a huge number of applications and occurrences in biology, physics, arts, computing, etc.. It is an infinite sequence and, below, only the first fourteen terms are represented

| i | 0 | 1 | 2 | 3 | 4 | 5 | 6 | 7 | 8 | 9 | 10 | 11 | 12 | 13 | 14 | … |
|---|---|---|---|---|---|---|---|---|---|---|----|----|----|----|----|---|
| F(i) | 0 | 1 | 1 | 2 | 3 | 5 | 8 | 13 | 21 | 34 | 55 | 89 | 144 | 233 | 377 | … |

Mathematically, it is defined by the formula $F(i)=F(i-2)+F(i-1)$ with seeds (or initial conditions) $F(0)=0$ and $F(1)=1$. Let us begin by making a connection with a result found in the second section above, following Eq.(8). The total number of atoms in the four ribonucleotides UMP, CMP, AMP and GMP is equal to 144 and this number could be understood as $(12+13+15+16)+4\times22$ that is the number of atoms in the four nucleobases U, C, A and G, on the one hand, and the number of atoms in the four invariant parts ribose+phosphate group (minus two water molecules), on the other (see above). Therefore, we have $144=56+88$. This partition could be derived in two ways, in connection to Fibonacci numbers. First, the sum of the Fibonacci numbers from $i=0$ to $i=12$ with *alternating signs* gives, in the last step, $-56+144=88$ or $56-144=-88$, depending on the sign we begin with (- or +). In both cases we are left with $144=56+88$. Secondly, we use Lucas's formula (1876), linking the Fibonacci numbers with the binomial coefficients (and also with Pascal's triangle)

$$F(n) \equiv \sum_{k=0}^{\lfloor (n-1)/2 \rfloor} C_k^{n-k-1} \qquad (17)$$

In this formula, $C_k^{n-k-1}$ is a binomial coefficient and $\lfloor x \rfloor$ denotes the floor function, that is the greatest integer less than or equal to $x$. In our case, we have for the 12$^{th}$ Fibonacci number $144=55+89$. (Observe that this not too far from $56+88$.) Using Lucas's relation for 55 and 89, we have $55 = C_0^9+C_1^8 + C_2^7 + C_3^6 + C_4^5 =1+8+21+20+5$ and $89=C_0^{10}+C_1^9+C_2^8 + C_3^7 + C_4^6 + C_5^5=1+9+28+35+15+1$. It remains only to group the *even* binomials together, with sum 56, and the *odd* binomials together, with sum 88, to get the desired result $144=56+88$.

Now, we turn to the main subject of this section and introduce the following two formulae for finite sums of Fibonacci numbers

$$S^{(n)} \equiv \sum_{i=0}^{n} F(i) = F(n+2) - 1 \qquad (18)$$

$$S^{(a,n)} \equiv \sum_{i=a}^{n} F(i) = F(n+2) - F(a+1) \qquad (19)$$

Note that the later reduces to the former, taking *a*=0. We consider now the following four interesting sums

$S^{(9)}=88$ (20)
$S^{(12)}=376=2\times188$ (21)
$S^{(5,14)}=979$ (22)
$S^{(4,14)}=982$ (23)

Here $S^{(9)}$ and $S^{(12)}$ are the same like $S^{(1,9)}$ and $S^{(1,12)}$, respectively. We shall see below that, from these four selected sums, we could derive several pertaining results concerning the nucleon-number distribution in various partitions of the amino acids. First of all, the second sum, from the above Fibonacci sequence, could be written as two *equal* parts: the sum of the *even* Fibonacci numbers $S_e^{(12)}=188$ and the sum of the *odd* Fibonacci numbers $S_o^{(12)}=188$. It is precisely here that we could see the first link to the number of nucleons in certain amino acids.

| class | Amino acid | #atoms | #nucleons |
|---|---|---|---|
| V | Proline | 8 (17) | 41 |
|   | Alanine (A) | 4 (13) | 15 |
|   | Threonine (T) | 8 (17) | 45 |
|   | Valine (V) | 10 (19) | 43 |
|   | Glycine (G) | 1 (10) | 1 |
| VI | Serine (S) | 5 (14) | 31 |
|   | Leucine (L) | 13 (22) | 57 |
|   | Arginine (R) | 17 (26) | 100 |
| II | Phenylalanine (F) | 14 (23) | 91 |
|   | Tyrosine (Y) | 15 (24) | 107 |
|   | Cysteine (C) | 5 (14) | 47 |
|   | Hystidine (H) | 11 (20) | 81 |
|   | Glutamine (Q) | 11 (20) | 72 |
|   | Asparagine (N) | 8 (17) | 58 |
|   | Lysine (K) | 15 (24) | 72 |
|   | Aspartic Acid (D) | 7 (16) | 59 |
|   | Glutamic Acid (E) | 10 (19) | 73 |
| III | Isoleucine (I) | 13 (22) | 57 |
| I | Methionine (M) | 11 (20) | 75 |
|   | Tryptophane (W) | 18 (27) | 130 |

**Table 3**: Atom and nucleon numbers in the 20 amino acids

As a matter of fact, each one of the three sextets leucine, serine and arginine occurs two times: one time as a quartet (four codons) and one time as a doublet (two codons). Knowing that the corresponding number of nucleons in the side chains of these amino acids are 57, 31 and 100, respectively (see Table 3) – with total sum 188 – we make therefore a connection with Eq.(21). Below, we shall make this connnection more tight.

Let us now form the following interesting linear combination of the four sums

$$S^{(9)}+S^{(12)}+2\times S^{(5,14)}+S^{(4,14)}=3404 \qquad (24)$$

3404 is equal to the number of nucleons in the 61 amino acids' side chains. As a matter of fact, from Table 3, there are 145 nucleons in (degeneracy) class IV, 188 nucleons in class VI, 660 nucleons in class II, 57 nucleons in class III and finally 205 nucleons in class I, so that $4\times145+6\times188+2\times660+3\times57+(75+130)=3404$. Strikingly, even small portions of the above sum still have interesting meaning. The first of them has already been mentioned above, in connection with the three sextets, and concerns $S^{(12)}=S_e^{(12)}+S_o^{(12)}=188+188$ to which we return now to add something also significant. As a matter of fact, the $B_0$-function of the sum $S^{(5,14)}=979$ (which occurs *two times* in Eq.(24)) leads to

$$B_0(979)=B_0(11\times89)=a_0(11\times89)+[SPI(11\times89)+\Omega(11\times89)] \qquad (25)$$

$$=100+[29+2]=100+31=131$$

and these last two numbers, 100 and 31, fit the number of nucleons nicely in the side chains of arginine and serine, respectively (see Table 3). Subtracting these two numbers from the sum $S^{(12)}$ above and rearranging we get finally $376=2\times(100+31+57)$, where we have used the property $S^{(12)}=S_e^{(12)}+S_o^{(12)}$, mentioned above The sum $S^{(12)}$, thanks to the $B_0$ function of the sum $S^{(5,14)}$, gives therefore *twice* the correct nucleon numbers of the three sextets arginine (100), serine (31) and leucine (57), respectively. The following other portion $S^{(9)}+S^{(5,14)}=1067$ gives the number of nucleons in the 17 amino acids in the classes I, II, III and IV (205+660+57+145=1067), (see Table 3). Note that by adding this last value to $S^{(12)}$, that is, $S^{(9)}+S^{(5,14)}+S^{(12)}$ gives $1067+2\times188=1443$, *i.e.*, the nucleon numbers in the 23 amino acids (counting the sextets two times). This is also a well known number. Finally, the last portion, $S^{(5,14)}+S^{(4,14)}=979+982=1961$, gives the number of nucleons in the remaining 38 degenerate amino acids (61-23=38): $145\times3+660\times1+188\times4+57\times2=1961$ and we have for the 61 amino acids, as given in one relation, in Eq.(24) above:

$$[S^{(9)}+S^{(5,14)}+S^{(12)}]+[S^{(5,14)}+S^{(4,14)}]=1443+1961=3404 \qquad (26)$$

We could also make another grouping: $1067+S_e^{(12)}=1067+188=1255$ and $1961+S_o^{(12)}=2149$ where 1255 is the number of nucleons in the 20 amino acids (see Table 3) and 2149 is the number of nucleons in the 41 degeneracy cases (here, in 1255, the sextets are counted only one time). Other interesting results could still be obtained. For example, we have $S^{(5,14)}+B_0(S^{(5,14)})=979+131=1110$ and, subtracting from the number 1443 above and, rearranging, we obtain $1443=333+1110$. These are also well known numbers, first obtained by Shcherbak (see Shcherbak, 2003), and correspond to the nucleon numbers in 8 four-codons and 15 non-four codons amino acids, respectively. The sum in Eq.(24) could also be used (*at second time*) to get the total number of nucleons in the 61 amino acids by including the blocks of the amino acids. As a matter of fact $2\times3404+S^{(5,14)}+B_0(S^{(5,14)})=3404+4514=7918$. The first number, 3404, as we have seen above, is the number of nucleons in the 61 amino acids' side chains and 4514 ($61\times74$) is the number of nucleons in the 61 (identical) blocks with 74 nucleons in each one of them.

As mentioned in the introduction, we could also establish the multiplet structure of the 20 amino acids and the total degeneracy. Let us take the following (new) sum of the first few Fibonacci numbers $S^{(1,6)}=1+1+2+3+5+8=20$. *This is the number of* (essential) *amino acids*. Now, among the six Fibonacci numbers, four already correspond to correct group-numbers: 1 (for example $F(2)$) for the triplet (I), 2 for the two singlets (M, W), 3 for the three sextets (L, S, R) and finally 5 for the five quartets (P, A, T, V, G). It suffices now to consider the two remaining Fibonacci numbers, $1=F(1)$, and $8=F(6)$, and add them to get the group number for the nine doublets $1+8=9$. This is in good agreement with a formula for the even-class amino acids established many years ago by Gavaudan (1971) that writes as $2^n+1$. Gavaudan showed that group-numbers corresponding to even-classes are in accordance with a geometrical progression when the class-numbers are inversely ordered by an arithmetic progression:

| Even classes | 2 | 4 | 6 |
|---|---|---|---|
| Corresponding groups | $2^3+1=9$ | $2^2+1=5$ | $2+1=3$ |

Our result above, $2^3+1=1+8=9$, corresponds therefore to the nine doublets (F, Y, C, H, Q, N, K, D, E). The sum $S^{(1,6)}$ above could also be used to get the degeneracy. Adding the six ranking indices $i$ together gives 21 so that the sum of the six Fibonacci numbers *and* their indices gives $S'=S^{(1,6)}+21=20+21=41$ which is the correct total degeneracy. We have therefore obtained the well known pattern in the form $S^{(1,6)}+S'=20+41=61$: 20 for the number of (essential) amino acids and 41 degenerate (codons) amino acids. We could even take $S^{(0,6)}$, in place of $S^{(1,6)}$, in the sum $S'$ above. This changes nothing except for the presence of two zeros, one in $S^{(0,6)}$ and one as its index, and we have now $S$'=(**0**+1+2+3+5+9)+(**0**+1+2+3+4+5+6)=41. A simple sorting of these numbers leads to an improved (better) result:

| group numbers | 1 | 2 | 3 | 5 | 1+8=9 | 20 |
|---|---|---|---|---|---|---|
| #degenerate codons | 2 | **0**+**0**=0 | 1+2+3+4+5=15 | 1+3+5+6=15 | 1+8=9 | 41 |
| total number of codons | 3 | 2 | 18 | 20 | 18 | 61 |

Therefore, the two zeros prove to be suitable for the two singlets as these latter have zero degeneracy.

Let us finally make a remark on all the sums of Fibonacci numbers used above. We considered $S^{(9)}$, $S^{(12)}$, *two times* $S^{(5,14)}$, $S^{(4,14)}$ and *two times* $S^{(1,6)}$ (see above). It is interesting that the total number of Fibonacci numbers, used in these sums, is equal to 64 which is the total number of codons. By separating the numbers of the sums, which appear two times, from the others, we get (9+12+11)+(2×10+2×6)=32+32=64 a pattern that we met in section 2, in connection with the pattern "2×3456". We could also separate the repeating numbers from the non-repeating ones and get (9+12+11+10+6)+(10+6)=48+16=64, which is just the first form of Eq.(10) in section 2: $\varphi(144)+\varphi(48)$=48+16=64. (In this latter reasoning, there are two possible cases leading to the same result: (i) either we simply disregard the above "improvement" which includes two supplementary zeros and use *two times* $S^{(1,6)}$ or (ii) we consider it by using *two times* $S^{(0,6)}$ and, in this case, we count *only* the non-zero Fibonacci numbers.)

At the end of this section, let us return to the pattern "2×3456" considered in the second section, and mentioned above, which could also be meant as 3456=3456. We know from the preceding section that $B_0[\sigma(3456)]$=61, that is Eq.(15)+Eq.(15)', so that the pattern "3456+3456" is equivalent to 61=61 or 61-61=0. We call this last relation a "*constraint equation*", and it will prove to be very useful, below, to derive several new patterns. There are four interesting cases.

Consider, first, the number of atoms in the precursors of the amino acids, as found in Eq.(12): $\sigma(144)$=403. It is known that the sum of all the divisors of a number has two parts: the sum of the proper divisors, i.e., $\sigma(n)-n$ the so-called aliquot sum, on the one hand, and the number $n$ itself. Applying this property to $n$=144, we get $\sigma(144)$=259+144. It remains to introduce the "*constraint equation*" -61+61=0, mentioned above, to get 403=198+205. From the data given in footnote 2, we have that there are 205 atoms in the 8 amino acid precursors *containing* phosphate: 25+4×18+3×36=205 ({H}, {G, **S**, C]; {W, F, Y}) and 198 atoms in the 15 amino acids precursors *not containing* phosphate 4×10+5×16+6×13=198 ({A, **L**, V}, {P, E, Q, **R**}, { T, M, I, D, N, K }). This partition into "phospho" and "non-phospho" amino acid precursors has been studied by (Rakočević and Jokić, 1996) and also considered by ourselves (Négadi, 2011). Our second case concerns the nucleon numbers in the 20 amino acids in the two classes of AARSs. There are 1316 nucleons in the 10 amino acids in class-II and 2088 nucleons in the 10 amino acids in class-I (see Table 2). We have seen above that $1067+S_e^{(12)}=1255$ (total nucleons in 20 amino acids side chains) and $1961+S_o^{(12)}=2149$ (total nucleons in 41 degenerate amino acids), with total sum 3404. Introducing the "*constraint equation*" in the expressions above, it gives 1255+61=1316 and 2149-61=2088 which is the desired result for the two classes. The third case is connected to the preceding one, where we add the blocks of the amino acids now. We found above 3404+4514=7918, the total number of nucleons in the 61 amino acids (side chains: 3404, blocks: 4514). Let us arrange the various Fibonacci sums in Eqs.(20)-(23) entering in the composition of the numbers 3404 and 4514 (see above) the following two parts

$$3\times S^{(5,14)}+S^{(4,14)}+S^{(12)} =4295 \tag{27}$$

and

$$2\times S^{(5,14)}+S^{(4,14)}+S^{(12)}+2\times S^{(9)}=3623 \tag{28}$$

Of course, the sum is still 7918. Now, the total number of nucleons in the class-I AARS (29 codons) is equal to 2088+29×74=4234 and the number of nucleons in the class-II AARS (32 codons) is equal to 1316+32×74=3684. (We mentioned already that the number of nucleons in the identical blocks is equal to 74.) It is now easy to get these last two numbers 4234 and 3684 from the Eqs.(27) and (28) for the two classes of AARSs. There is sufficient to introduce our "*constraint equation*" in their sum: (4295-61)+(3623+61)=4234+3684.

## 4. CONCLUDING REMARKS

In this work, we considered the total number of atoms in the four ribonucleotides *uridine monophosphate* (UMP), *cytidine monophosphate* (CMP), *adenosine monophosphate* (AMP) and *guanosine monophosphate* (GMP), 144, as a *key parameter*, to start with. In the first section, using Euler's $\varphi$-function (and also some other few arithmetic functions as the sum of divisors $\sigma$-function and the $B_0$-function, see text), we derived from the above parameter several interesting quantities, among them and mainly, a mathematical relation describing the "*condensation*" of the three basic sub-units a nucleobase, a ribose and a phosphate group, to form a ribonucleotide, the total number of *atoms* in the 64 codons of the genetic code table as well as an *invariance* equation for this number. Also, we established, mathematically, Rakočević's pattern "2×3456" for the total number of atoms in the 64 codons, and showed its occurring in other recent works on the genetic code. In the second section, we used a few selected sums of Fibonacci numbers to derive the number of *nucleons* in the amino acids, in various characteristic patterns, and also the multiplet pattern of the 20 essential amino acids as well as the degeneracy. In summary, Euler's $\varphi$-function has been shown to be associated to the atom numbers in the 64 codons and a combination of selected finite sums of Fibonacci numbers has been shown to be associated to the nucleon numbers in the amino acids (degeneracy included).

Remarkably, the number 144 – which is the number of atoms in the four ribonucleotides and our starting point in this work – is also a *distinguished* Fibonacci number, the 12$^{th}$. As a matter of fact, it has been shown some fifty years ago (Cohn, 1964) that 0, 1 and 144 are the only *perfect squares* and for 144, in particular, one has $F(12)=12^2$. Furthermore, this number is associated to another fundamental ingredient of life, *water*: it is equal to the total number of distinguishable configurations of two neighboring water molecules in the "body-centered cubic lattice-gas model of water" (Besseling and Lyklema, 1994). In that model, each water molecule has 12 possible orientations and for two (neighboring) water molecules there are $12^2$ possible configurations.

## REFERENCES

Besseling NAM and Lyklema J. (1994) Equilibrium Properties of Water and Its Liquid-Vapor Interface. *Phys Chem*; 98: 11610-11622.
*Cohn, J. H. E. (1964)* On square Fibonacci numbers, *J. London Math. Soc.,* 537–540.
Downes, A.M., Richardson B.J. (2002) Relationships between genomic base content distribution of mass in coded proteins *J Mol Evol* , 55, 476-490.
Gavaudan, P. (1971) The genetic code and the origin of life. In *Chemical Evolution and the origin of life*; Buvet R and Ponnamperuma C. Eds. North-Holland Publishing Company, 432-445.
Gilbert, G. (1986) The RNA World, *Nature*, 319:618.
Négadi, T. (2001) Symmetry and proportion in the genetic code, and genetic information from the basic units of life. *Symmetry: Culture and Science,* 12 (3-4), 371-393.
Négadi, T. (2005) Symmetry and Information in the genetic code, in *Advances in Bioinformatics and its Applications*, World Scientific Publishing and Co. (2005).
Négadi, T. (2007) The genetic code multiplet structure, in one number, *Symmetry: Culture and Science*, 18 (2-3), 149-160.
Négadi, T. (2008) The genetic code via Gödel encoding, *The Open Physical Chemistry Journal*, 2, 1-5.
Négadi T. (2009) The genetic code degeneracy and the amino acids chemical composition are connected, *NeuroQuantology*, 7, 1, 181-187 (Available also at http://arxiv.org: q-bio. OT/0903.4131).
Négadi, T. (2011) On Rakočević's  amino acids biosynthetic precursors relations, *NeuroQuantology*, 9, 4.

Négadi, T. (2012) The irregular (integer) tetrahedron as a warehouse of biological information, *Symmetry: Culture and Science*, 23 (3-4), 403-426.( available also at: arxiv.org/pdf/1207.3454.)

Nirenberg, M., Leder, P., Bernfield, M., Brimacombe, R., Trupin, J., Rottman, F., and O'Neal, C. (1965) NA codewords and protein synthesis, VII. On the general nature of the RNA code. *Proc.Natl. Acad. Sci. U. S. A.* 53 (5), 1161–1168.

Perez, J.-C. (2010) Codon populations in single-stranded whole human genome DNA are fractal and fine tuned by the golden ratio 1.618, *Interdiscip Sci Comput Life Sci,* 2, 1–13.

Petoukhov, S.V. (1012) Symmetries of the genetic code, hypercomplex numbers and genetic matrices with internal complementarities, *Symmetry: Culture and Science*, 23, Nos.3-4, 275-301.

Rakočević, M. M. (1997) Two classes of the aminoacyl-tRNA synthetases in correspondence with the codon path cube, *Bull. Math. Biol.*, 59, 645-648.

Rakočević,M. M. (1997) *Genetic code as a unique system*, SKC, Niš.

Rakočević, M. M. (2006) Genetic code as a harmonic system, *arXiv:qbio/0610044v1* [q-bio.OT].

Rakočević, M.M. (2011) Genetic code: four diversity types of protein amino acids, *arXiv:1107.1998v2* [qbio.OT].

Rakočević, M.M. (2009) Genetic Code Table: A note on the three splittings into amino acid classes, *arXiv:qbio/ 0903.4110v1*[q-bio.BM].

Rakočević, M.M. (2013) Harmonic mean as a determinant of the genetic code, *arXiv:qbio/1305.5103v1* [q-bio.OT].

Rakočević MM and Jokić A. (1996) Four stereochemical types of protein amino acids: synchronic determination with chemical characteristics, atom and nucleon number. *J. Theor. Biol.* 183, 345–349.

Shcherbak, V.I. (2003) Arithmetic inside the genetic code. *BioSystems*, 70, 187-209.

Shcherbak, V.I. (1994) Sixty-four triplets and 20 canonical amino acids of the genetic code: the arithmetical regularities. Part II. *J. Theor. Biol.* 166, 475-477.

Yang, C. M. (2004) On the 28-gon symmetry inherent in the genetic code intertwined with aminoacyl-tRNA synthetases--the Lucas series. *Bull. Math. Biol.* 66, 1241-1257.